\PassOptionsToPackage{svgnames}{xcolor}
\documentclass[twocolumn,groupedaddress,aps,preprintnumbers,amsmath,amssymb,sort&compress,nofootinbib,prd]{revtex4-2}
\usepackage{xcolor,ulem}

\usepackage{amsfonts,amsmath,amssymb,ascmac,bm,tensor,physics}
\usepackage{graphicx}
\usepackage{dcolumn}
\usepackage{comment}
\usepackage{slashed}
\usepackage{color}
\usepackage[mathscr]{eucal}
\usepackage[utf8]{inputenc}
\usepackage{float}
\usepackage[nohyperlinks,nolist]{acronym}
 
\usepackage{cancel}

\usepackage[allcolors=black]{hyperref}
\usepackage[capitalize]{cleveref}
\newcommand{\para}{\parallel}
\newcommand{\bfx}{{\mathbf{x}}}
\newcommand{\bfv}{{\mathbf{v}}}
\newcommand{\bfk}{{\mathbf{k}}}
\def\vev#1{{\left\langle #1 \right\rangle}}

\newcommand{\Mpc}{\mathrm{Mpc}}

\long\def\comment#1{}

\def\vev#1{{\left\langle #1 \right \rangle}}

\begin{document}


\title{Peculiar-velocity distribution functions and 21-cm fluctuations}
\author{Ryan Yuran Zhang}
\author{Marc Kamionkowski}
\affiliation{William H.\ Miller III Department of Physics and
Astronomy, Johns Hopkins University, 3400 N. Charles Street,
Baltimore, MD 21218, U.S.A.}

\begin{abstract}
Predictions for observables involving the cosmological 21-cm background require calculations of spatial correlations of star formation rate densities (SFRDs) which have a nonlinear dependence on the baryon-dark matter relative velocity.  Prior work derived these SFRD correlations with a simplifying assumption that neglected the difference between the correlations of the components of the velocity parallel and perpendicular to the separation between the two points being correlated.  Here we calculate the full joint PDF of the squares of the peculiar velocity at two different points.  The error that arises in predictions for 21-cm fluctuations if this subtlety is overlooked is  generally less than a few percent, but it can be larger for some values of wavenumber $k$ and redshift $z$ if there are cancellations between different contributions to the total signal.  The correct expression is easily implemented and increases the run time of the code by only a few percent. 
\end{abstract}

\date{\today}
\maketitle

\section{Introduction}

As the statistics in galaxy surveys and \ac{cmb} maps improve, correlation functions that are higher order in cosmological perturbations begin to become accessible.  Among the many possible such correlations, there are now several being studied that involve the square of the peculiar-velocity field.  For example, 21-cm intensity fluctuations may exhibit \acp{vao}~\cite{Munoz:2019fkt,Munoz:2019rhi}  that arise from correlations between observables proportional to the square of the peculiar velocity field. Similarly, observables proportional to the velocity squared have been suggested for \ac{ksz} tomography~\cite{Smith:2016lnt,AnilKumar:2025tbe,AnilKumar:2025wyt}. 

In linear perturbation theory, the peculiar velocity is obtained from the gradient of the density field which is itself a realization of a Gaussian random field.  Each of the three components $v_i$ (for $i=x,y,z$) of the peculiar velocity is thus a Gaussian random variable, and the one-point \ac{pdf} of $\bfv^2=v_x^2+v_y^2+v_z^2$, the magnitude squared, is thus a $\chi^2_n$ distributed variable with $n=3$ degrees of freedom (the Maxwell-Boltzmann distribution).  It is then natural to
surmise that the joint \ac{pdf} for the values of $\bfv^2$ at two different points is that for two correlated $\chi_n^2$ degrees of freedom, related to the Wishart distribution, as shown, for example, in the Appendix of Ref.~\cite{Givans:2023kbg} (see also Refs.~\cite{Percival:2006ss,Hamimeche:2008ai}).  This assumption was made in Ref.~\cite{Cruz_2025} (and implemented in {\tt Zeus21}\footnote{\tt https://github.com/JulianBMunoz/Zeus21}) to obtain correlation functions for \acp{sfrd}, which depend nonlinearly on the square of the peculiar velocity.  

However, the joint \ac{pdf} for the values of $\bfv^2$ at two
different points is not strictly speaking described in this way, as the correlations of the components of the velocity aligned with the separation of the two points being correlated differ from those for the components transverse to that separation.
Here we explain and quantify this subtlety and explore its implications for calculations of 21-cm observables.  Although the impact on observables is relatively small for most wavenumbers and redshifts, it can be larger for wavenumbers and redshifts where there are cancellations in the different contributions to the total signal.  The correct result is easily implemented in {\tt Zeus21}.

The plan for this paper is as follows:  In the next
Section, we review the calculation of the correlations of components of
velocities aligned with and transverse to the separation between
the two points being correlated.  In Section \ref{sec:jointpdf}, we describe the complete joint \ac{pdf}.  In Section \ref{sec:21cm} we discuss the specific SFRD correlation required by {\tt Zeus21}, derive the exact result, and compare it with the approximation used in Ref.~\cite{Cruz_2025}.  We then show in Section \ref{sec:observables} how our calculation impacts 21-cm observables, and we conclude in Section \ref{sec:conclusions}.  An Appendix calculates the complete joint \ac{pdf} and shows how it differs from the simple approximation adopted in Ref.~\cite{Cruz_2025}.  A second appendix discusses the implementation  in {\tt Zeus21}.

\section{The velocity two-point correlation function}
\label{sec:velocitycorrelations}

We start with a cosmological density perturbation $\delta(\bfx)$, which is assumed to be a realization of a Gaussian random field with power spectrum $P(k)$.  This means that the Fourier amplitudes
\begin{equation}
     \tilde \delta(\bfk) = \int\, \dd[3]{x}\, e^{i\bfk \cdot
     \bfx} \delta(\bfx),
\label{eqn:FT}
\end{equation}
satisfy
\begin{equation}
     \vev{ \tilde\delta(\bfk) \tilde \delta^*(\bfk') } =
     (2\pi)^3 \delta_D(\bfk-\bfk') P(k),
\label{eqn:Pkdefn}
\end{equation}
where the angle brackets denote an average over all
realizations, and $\delta_D(\bfk-\bfk')$ is the Dirac delta
function. 

At linear order, a velocity field $\bfv(\bfx)$ must be $\bfv(\bfx) \propto \nabla \delta(\bfx)$.  For the 21-cm calculation the relevant velocity field is the baryon-dark matter relative velocity (we leave out the usual subscript ``bc'' to reduce notational clutter).  Its Fourier components $\tilde \bfv(\bfk)$ are therefore related to those of the density field through
\begin{equation}
    \tilde\bfv(\bfk) = i f(k) \bfk \tilde \delta(\bfk),
\label{eqn:vdeltarelation}
\end{equation}
where $f(k)$ is a transfer function.

\begin{figure}
\includegraphics[width=\columnwidth]{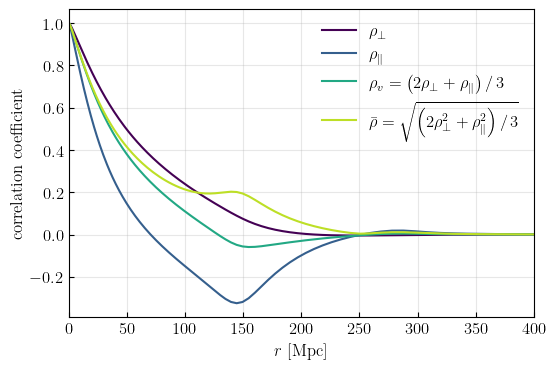}
\caption{The velocity correlation coefficients as functions of the separation $r$ in units of $\Mpc$. We show $\rho_\para(r)$ and $\rho_\perp(r)$ for, respectively, the components aligned with and transverse to the separation between the two points at which the velocity is correlated.  We also show the isotropic correlation coefficient $\rho_v = (\rho_\parallel(r)+2 \rho_\perp(r))/3$, as well as the quantity, $\left[ \left(\rho_\parallel^2(r) + 2\rho_\perp^2(r) \right)/3 \right]^{1/2}$ that appears in the approximation, given here in Eq.~(\protect\ref{eqn:cruzresult}), of $\xi_U(r)$ in Ref.\protect\cite{Cruz_2025}.}
\label{fig:rho}
\end{figure}

We now calculate the two-point correlation function for the
components of the velocity field aligned with the separation
between the two points at which the velocity is correlated.  To
do so, let us take one point as the origin and the second as
 the $\hat z$-axis at a distance $r$.  Using
Eqs.~(\ref{eqn:FT}), (\ref{eqn:vdeltarelation}), and
(\ref{eqn:Pkdefn}), the correlation of the components aligned
with the separation is then,
\begin{eqnarray}
     \xi_\parallel(r) &\equiv& \vev{ v_z(\mathbf{0})v_z(r\hat z) }
     \nonumber \\
     & = &  \int
     \frac{\dd[3]{k}}{(2\pi)^3} \frac{\mu^2}{k^2} e^{-i k r \mu}\, P(k)\left[f(k)\right]^2
     \nonumber \\
     &=& \frac{1}{4\pi^2} \int_0^\infty\, \dd{k}\, P(k) \left[f(k)\right]^2
     \int_{-1}^1 \, \dd{\mu}\, \mu^2 e^{-ikr\mu} \nonumber \\
     &=& \frac{1}{2\pi^2} \int_0^\infty\, \dd{k}\, P(k)\left[f(k)\right]^2
     \left[j_0(kr) - \frac{2}{kr} j_1(kr) \right].\nonumber \\
\label{eqn:parallel}
\end{eqnarray}
The correlation of the components perpendicular is
\begin{eqnarray}
     \xi_\perp(r) &\equiv& \vev{ v_x(\mathbf{0})v_x(r\hat z)
     }\nonumber \\
     &=& \frac12 \frac{1}{4\pi^2}
     \int_0^\infty\, dk\, P(k) \left[f(k)\right]^2 \int_{-1}^1 \, d\mu\,
     (1-\mu^2) e^{-ikr\mu} \nonumber \\
     &=& \frac12 \frac{1}{\pi^2} \int_0^\infty\, \dd{k}\,
     P(k)\left[f(k)\right]^2 \frac{j_1(kr)}{kr},
\end{eqnarray}     
and similarly $\vev{ v_y(\mathbf{0})v_y(r\hat z) } = \xi_\perp(r)$.
The correlation for the dot product then comes out to
\begin{eqnarray}
     \xi_v(r) &\equiv & \vev{ \bfv(\mathbf{0})\cdot \bfv(r\hat z) }
     \nonumber \\
     &=& \frac{1}{2\pi^2} \int_0^\infty\, \dd{k}\, P(k)\left[f(k)\right]^2
     j_0(kr),
\label{eqn:dotcorrelation}     
\end{eqnarray}    
as often seen in the literature.  In these expressions, $j_n(x)$
are the spherical Bessel functions with $j_0(x) = \sin(x)/x$ and
$j_1(x) = (\sin x - x \cos x)/x^2$.  \cref{fig:rho} shows the correlation coefficients $\rho_\perp(r)\equiv\xi_\perp(r)/\bar v^2$, $\rho_\para(r)\equiv\xi_\para(r)/\bar v^2$, and $\rho_v(r)\equiv\xi_v(r)/(3\bar v^2)=[\rho_\parallel(r) + 2 \rho_\perp(r)]/3$.  Here $\bar v^2 = \xi_\parallel(0)=\xi_\perp(0)=\xi_v(0)/3$.

\section{The joint PDF for a $\chi^2_{n=3}$ distribution and for the velocity squared}
\label{sec:jointpdf}

We now determine the joint \ac{pdf} for the values $v_1^2=|\bfv_1|^2$ and $v_2^2=|\bfv_2|^2$ of the squares of the velocities $\bfv_1$ and $\bfv_2$ at two points separated by a distance $r$.  As before, without loss of generality, we take the two points to be separated in the $\hat z$-direction.  The six components (i.e., $v_{1x}$, $v_{1y}$, $v_{1z}$, $v_{2x}$, $v_{2y}$, $v_{2z}$) of these two velocity vectors are drawn from a Gaussian distribution with variances $\bar v^2$.  The only nonzero covariances are $\vev{ v_{1x} v_{2x}} =\vev{v_{1y} v_{2y}} = \rho_\perp \bar v^2$ and $\vev{v_{1z} v_{2z}} = \rho_\parallel\bar v^2$.

The joint \ac{pdf} for the components of $\bfv_1$ and $\bfv_2$ is then
\begin{eqnarray}
     P_v\left(\bfv_1,\bfv_2 \right) &=& \frac{1}{N}
    \exp\left[-\frac{1}{2\bar v^2}\left(\frac{v_{1z}^2+v_{2z}^2-2\rho_{\parallel}v_{1z}v_{2z}}{1-\rho_{\parallel}^2} \right. \right. \nonumber \\  & & \left. \left. +\frac{v_{1\perp}^2+v_{2\perp}^2-2\rho_{\perp} \bfv_{1\perp}\cdot \bfv_{2\perp}}{1-\rho_{\perp}^2} \right) \right],
\end{eqnarray}
with
\begin{equation}
    N= (2\pi \bar v^2)^3 \left( 1- \rho_\parallel^2 \right)^{1/2} \left( 1 -\rho_\perp^2 \right).
\label{eqn:Neqn}    
\end{equation}

The joint \ac{pdf} for $X_1\equiv v_1^2$ and $X_2\equiv v_2^2$, is then
\begin{eqnarray}
     P(X_1,X_2) &=& \int\, \dd[3]{v_1} \, \int\, \dd[3]{v_2} P_v(\bfv_1,\bfv_2) \nonumber \\
    & &  \times \, \delta_D(X_1-v_1^2) \delta_D(X_2-v_2^2).
\label{eqn:short}     
\end{eqnarray}
\cref{sec:appendix} shows how this is evaluated and related in the limit $\rho_\perp\to \rho_\parallel$ to the expression used in Ref.~\cite{Cruz_2025}.  However, we will now see that Eq.~(\ref{eqn:short}) is a good starting point for the required calculation.

For reference, Ref.~\cite{Cruz_2025} approximates this joint \ac{pdf} with Eq.~(\ref{eqn:short}) but using a joint velocity \ac{pdf} where $\rho_\perp$ and $\rho_\parallel$ are both taken to be $\bar\rho$.

\section{Implications for 21-cm calculation}
\label{sec:21cm}

We now show how these \acp{pdf} enter into {\tt Zeus21}.  The correlations required there are for exponential fields, $e^{\lambda_\alpha Y(\bfx)}$, where $Y=v^2(\bfx)/\bar v^2$, and  $\lambda_\alpha$ is a coefficient that depends on the specific quantity (e.g., stellar density, star formation rate, etc.). Most generally, the correlation will be a cross-correlation between two populations labeled by $\alpha=1$ and $\alpha=2$, in which case we need cross-correlation functions for the quantities $U_1 =  e^{-\lambda_1 Y_1(\bfx_1)}$ and $U_2 = e^{-\lambda_2 Y_2(\bfx_2)}$ at two different points separated by a comoving distance $r=|\bfx_1-\bfx_2|$.  (Note that our $Y_\alpha$ is referred to as $\tilde\eta_\alpha$ in Ref.~\cite{Cruz_2025}, and our $\bar v^2$ is $\sigma^2$ there.)

The correlation function for $Y$ is
\begin{equation}
    \xi_Y(r) \equiv \frac{ \vev{ Y(\bfx_1) Y(\bfx_2)}-\vev{Y}^2}{\vev{Y}^2}
    =\frac{2\rho_\parallel^2+4\rho_\perp^2}{9}.
\end{equation}

The required two-point correlation for $U_\alpha$ is then
\begin{equation}
    \xi_U(r) = \frac{\vev{U_1 U_2}-\vev{U_1}\vev{U_2}}{\vev{U_1}\vev{U_2}}.
\label{eq:eta_U}
\end{equation}
The expectation values of $U_\alpha$ are
\begin{equation}
    \vev{U_\alpha} \equiv\ \qty(1+2\lambda_\alpha)^{-3/2},\label{eq:expUi}
\end{equation}
while that for the product is
\begin{align}
    \vev{U_1 U_2}\ \equiv&\ \iint U_1 U_2 P\qty(Y_1,Y_2)\dd{Y_1}\dd{Y_2}\ \\
    =&\ \iint e^{-\qty(\lambda_1 Y_1+\lambda_2 Y_2)}P\qty(Y_1, Y_2)\dd{Y_1}\dd{Y_2}.
\end{align}
Given the Dirac delta functions in Eq.~(\ref{eqn:short}), this becomes
\begin{equation}
     \vev{U_1 U_2} = \int\, \dd[3]{v_1} \, \int\, \dd[3]{v_2}\, e^{-\lambda_1 Y_1-\lambda_2 Y_2} P_v(\bfv_1,\bfv_2).
\end{equation}
The six integrals can now be separated as $\vev{U_1 U_2} = I_\parallel I_\perp^2$, with
\begin{equation}
     I_\parallel = \int \, \dd{v_1}\, \dd{v_2}\, e^{-(\lambda_1 v_1^2 +\lambda_2 v_2^2)/\bar v^2} P_\parallel(v_1,v_2),
\label{eqn:Iparallelintegral}
\end{equation}
and
\begin{equation}
     P_\parallel(v_1,v_2) = \frac{1}{2\pi \bar v^2} \exp\left[-\frac{1}{2\bar v^2}\left(\frac{v_{1}^2+v_{2}^2-2\rho_{\parallel}v_{1}v_{2}}{1-\rho_{\parallel}^2} \right) \right].
\end{equation}
Eq.~(\ref{eqn:Iparallelintegral}) then evaluates to
\begin{equation}
     I_\parallel = \left[1+2\lambda_1 + 2\lambda_2 + 4\lambda_1 \lambda_2 (1-\rho_\parallel^2) \right]^{-1/2}.
\end{equation}
The result for $I_\perp$ is the same with $\parallel\to\perp$.

Putting it all together, Eq.~(\ref{eq:eta_U}) becomes
\begin{equation}
    \xi_U = \left[ \left(1-K \rho_\parallel^2 \right) \left(1-K\rho_\perp^2 \right)^2 \right]^{-1/2}-1,
\label{eqn:newresult}
\end{equation}
where
\begin{equation}
 K \equiv \frac{4\lambda_1\lambda_2}{(1+2\lambda_1)(1+2\lambda_2)}.
\end{equation}

The new result in Eq.~(\ref{eqn:newresult}) is to be contrasted with that
\cite{Cruz_2025}
\begin{equation}
     \xi_U = \left[1-K(\rho_\parallel^2+ 2 \rho_\perp^2)/3\right]^{-3/2}-1,
\label{eqn:cruzresult}     
\end{equation}
obtained under the assumption that all three components of the peculiar velocity have the same correlation coefficient $\rho$.   Written in this way, it is clear that the new result in Eq.~(\ref{eqn:newresult}) reduces to the result of Ref.~\cite{Cruz_2025} in the limit $\rho_\perp^2 \to \rho_\parallel^2$. However, in general it differs.

\begin{figure}
\includegraphics[width=\columnwidth]{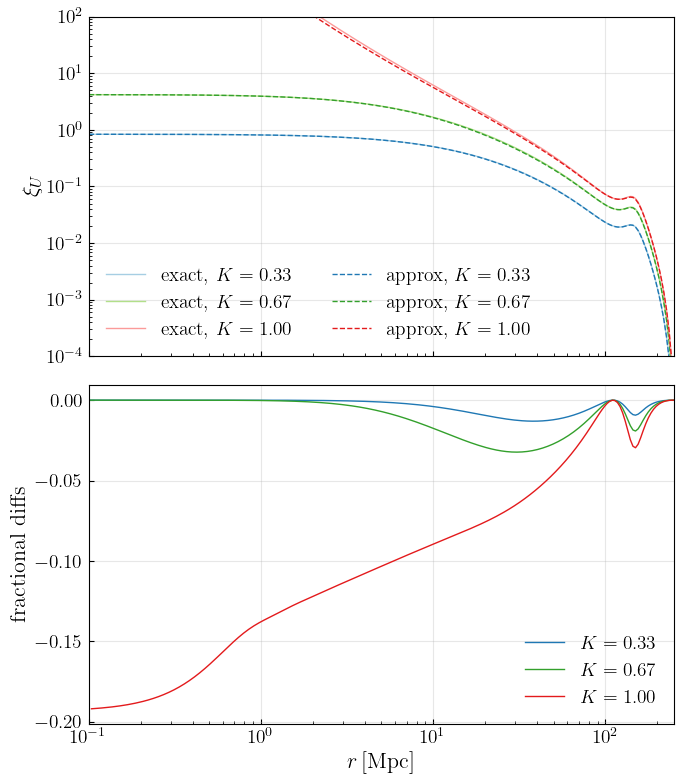}
\caption{Here we show $\xi_U(r)$ from the exact and approximate calculations, using the velocity correlation functions in Fig.~\protect\ref{fig:rho} for values of $K=0.33$, 0.67, and 1.  The values of $K$ required for 21-cm fluctuations are typically around $K\simeq0.5$.}
\label{fig:xiU}
\end{figure}

Most generally, $0<K<1$, but typical values of $\lambda_\alpha$ for the 21-cm calculation are $\simeq1$, implying $K\simeq0.5$ typically.  The correlations die off at large distances, and so as $r\to\infty$, $K\rho_\perp^2,K\rho_\parallel^2 \ll1$, and the correct correlation function becomes $\xi_U\simeq K (\rho_\parallel^2+ 2 \rho_\perp^2)/2$ and, in this limit, coincides with the approximation.  The correction is expected to be larger at smaller $r$, where $\rho_\parallel(r)$ and $\rho_\perp(r)$ become larger, but this is then counteracted by the fact that both approach unity as $r\to0$.  Fig.~\ref{fig:xiU} shows results for the fractional difference in $\xi_U(r)$ for several values of $K$.

\section{Effects of the anisotropy on the 21-cm power spectrum}
\label{sec:observables}

\begin{figure}[t]
\includegraphics[width=0.5\textwidth]{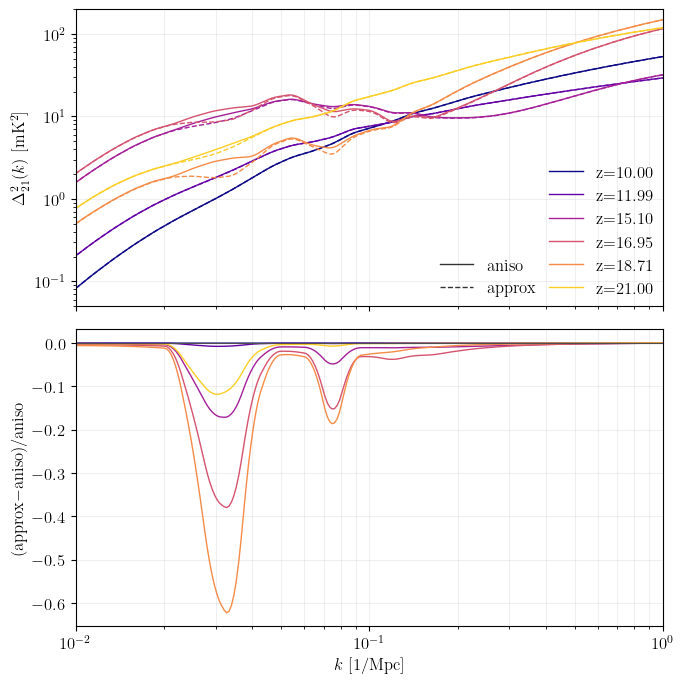}
\caption{The top panel shows the 21-cm intensity-fluctuation power spectrum $\Delta_{21}^2(k)$ for several redshifts.  The solid curves are the {\tt Zeus21} calculation, while the dashed are obtained by modifying {\tt Zeus21} to include the correct squared-velocity \acp{pdf}.  The calculations include Pop III stars and baryon-dark matter  relative velocities with the parameters used in the {\tt Zeus21} tutorial.  The correction is very small for most wavenumbers $k$ and redshifts $z$, but becomes large for a limited range of wavenumbers at redshifts $z\simeq15$ when there are cancellations between the different contributions to the total signal.
}
\label{fig:Delta}
\end{figure}

The correlation function $\xi_U(r)$ enters into expressions for correlation functions of \acp{sfrd}, which then contribute to 21-cm fluctuations.  However, these are not the whole story; there are other sources of 21-cm fluctuations.  Thus, to assess the impact of the exact \ac{pdf} calculation, we run {\tt Zeus21} using the exact and approximate expressions for $\xi_U(r)$.   The top panel of Fig.~\ref{fig:Delta} shows  the 21-cm power spectra $\Delta_{21}^2(k)$ as a function of wavenumber $k$ for several redshifts $z$ for the isotropic assumption in {\tt Zeus21} (solid curve) and the calculation described here (dashed).  The bottom panel shows the fractional differences between the approximate and exact calculation.  The differences are no more than a few percent for most wavenumbers and redshifts, but get larger for a limited range of $k$ at redshifts $z\simeq15$ where there are cancellations between different contributions to the total signal.  These calculations were performed for the default parameters in {\tt Zeus21}.  We note the precise magnitude of the correction, as well as the wavenumbers and redshifts where cancellations may occur, may be highly model dependent.

\section{Conclusions}
\label{sec:conclusions}

To conclude, we have discussed a subtlety in the calculation of the PDF of the squares of the peculiar velocity and investigated the implications for 21-cm fluctuations.  While the error introduced by overlooking this effect is generally small compared with other current sources of uncertainty, it may become big for wavenumbers and redshifts where there are cancellations between different contributions to the total signal.   It is easily implemented (as we detail in \cref{app:B} below) in the existing code and may become important in the future as we seek to interpret increasingly precise measurements.

\begin{acknowledgments}
We thank Hector Cruz for discussions.  RYZ acknowledges the support of a Johns Hopkins University Summer Provost's Undergraduate Research Award.
This work was supported by NSF Grant No.\ 2412361, NASA ATP Grant No.\ 80NSSC24K1226, and the Templeton Foundation.
\end{acknowledgments}

\appendix

\section{The full joint PDF}
\label{sec:appendix}

We evaluate the full joint \ac{pdf} by writing $\bfv_1$ and $\bfv_2$ in Eq.~(\ref{eqn:short}) in spherical coordinates taking, without loss of generality $v_1$ to lie in the $xz$-plane, with azimuthal coordinate $\phi=0$.  The joint PDF then becomes,
\begin{widetext}
\begin{eqnarray}
    P(X_1,X_2)
\comment{  &=&   \int_{-\infty}^{\infty} d{v_{1z}}\int_0^{2\pi}  d{\phi_1}\int_0^{\infty}v_{1\perp} d{v_{1\perp}} \int_{-\infty}^{\infty}  d{v_{2z}}\int_0^{2\pi}d{\phi_2}\int_0^{\infty}v_{2\perp} d{v_{2\perp}}\ P_v\left(\bfv_1,\bfv_2 \right) \delta \left(X_1-v_{1z}^2-v_{1\perp}^2 \right )\delta\left(Y_1-v_{2z}^2-v_{2\perp}^2 \right)\nonumber \\
    &=&2\pi \int_{-\infty}^{\infty}d{v_{1z}}\int_0^{\infty}\frac{1}{2}d{v_{1\perp}^2} \int_{-\infty}^{\infty} d{v_{2z}} \int_0^{2\pi} d{\phi_2}\int_0^{\infty}\frac{1}{2} d{v_{2\perp}^2} P_v\left(\bfv_1,\bfv_2 \right)\delta\left(X_1-v_{1z}^2-v_{1\perp}^2\right)\delta\left(X_2-v_{2z}^2-v_{2\perp}^2 \right) \nonumber\\
    &=&\frac{\pi}{2N} \int_{-\sqrt{X_1}}^{\sqrt{X_1}} d{v_{1z}}\int_{-\sqrt{X_2}}^{\sqrt{X_2}} d{v_{2z}}\int_0^{2\pi} d{\phi_2}\nonumber \\
    & &\ \ \ \ \  \times \exp\left[-\frac{1}{2\bar v^2}\left(\frac{v_{1z}^2+v_{2z}^2-2\rho_{\para}v_{1z}v_{2z}}{1-\rho_{\para}^2 }+\frac{X_1-v_{1z}^2+X_2-v_{2z}^2-2\rho_{\perp}\sqrt{X_1-v_{1z}^2}\sqrt{X_2-v_{2z}^2}\cos\phi_2}{1-\rho_{\perp}^2 }  \right)\right] \nonumber \\
}    
    &=&\frac{\pi}{N}\exp \left[-\frac{X_1+X_2}{2\bar v^2 \left(1-\rho_{\perp}^2 \right)}\right] \int_{-\sqrt{X_1}}^{\sqrt{X_1}} d{v_{1z}}\int_{-\sqrt{X_2}}^{\sqrt{X_2}} d{v_{2z}}\, I_0 \left(\frac{\rho_{\perp}\sqrt{X_1-v_{1z}^2}\sqrt{X_2-v_{2z}^2}}{\left(1-\rho_{\perp}^2 \right)\bar v^2} \right) \nonumber \\
    &  &\ \ \ \ \ \times \exp \left[-\frac{1}{2\bar v^2}\left(\frac{\rho_{\para}^2-\rho_{\perp}^2}{\left(1-\rho_{\para}^2 \right)\left(1-\rho_{\perp}^2 \right)} \left(v_{1z}^2+v_{2z}^2 \right)-\frac{2\rho_{\para}}{1-\rho_{\para}^2}v_{1z}v_{2z} \right) \right],
\end{eqnarray}
\end{widetext}
where $I_0(x)$ is a modified Bessel function of the first kind, and $N$ is given in Eq.~(\ref{eqn:Neqn}).

In the limit that $\rho_\perp \rightarrow \rho_\para \equiv\rho$, this becomes equivalent to the results in Refs.~\cite{Givans:2023kbg,Cruz_2025}; i.e.,
\begin{eqnarray}
    P_{\rm sym}(X_1,X_2) & = & \frac{(1-\rho^2)^{3/2} (X_1 X_2)^{1/2}}{4\pi} \exp\left(- \frac{X_1+X_2}{2(1-\rho^2)}\right) \nonumber \\ & & \times \int_{-1}^1 \exp\left( \mu \frac{\rho\sqrt{X_1 X_2}}{(1-\rho^2)}\right).
\end{eqnarray}


\section{Changes to {\tt Zeus21}}
\label{app:B}

The implementation of the corrected PDFs in {\tt Zeus21} is straightforward. A few changes are required in {\tt cosmology.py} and then more in {\tt correlations.py}.  The calculation of $P_\eta(k)$ in {\tt cosmology.py} must be replaced by calculations of $P_{\eta,\perp}(k)$ and $P_{\eta,\parallel}(k)$.  To get the former, the factor $6 \psi_0^2+ 3 \psi_2^2$ in the expression for $P_\eta(k)$ is replaced by $6(\psi_0-\psi_2/2)^2$ and the latter is obtained by replacing that factor with $6(\psi_0+\psi_2)^2$.

The quantity {\tt xiNumerator} in the function {\tt get\_xi\_Sum\_2expEta} in {\tt correlations.py} must then be altered so that the four evaluations of $\xi_U$ are corrected.  There are several steps in {\tt correlations.py} intermediate between the evaluation of $P_\eta(k)$ in {\tt cosmology.py} and its appearance in {\tt xiNumerator} that must be modified to replace $P_\eta(k)$ with $P_{\eta,\parallel}(k)$ and $P_{\eta,\perp}(k)$.

These modifications require two evaluations of {\tt mcfit} (which transforms between correlation functions and power spectra) to obtain $P_{\eta,\perp}(k)$ and $P_{\eta,\parallel}(k)$, rather than one for $P_\eta(k)$, but the increase to the total runtime is only a few percent.

\begin{acronym}
    \acro{pdf}[PDF]{probability distribution function}
    \acro{cmb}[CMB]{cosmic microwave background}
    \acro{vao}[VAO]{velocity acoustic oscillation}
    \acro{ksz}[kSZ]{kinematic Sunyaev-Zeldovich}
    \acro{sfrd}[SFRD]{star formation rate density}
    \acrodefplural{sfrd}[SFRDs]{star formation rate densities}
\end{acronym}

\bibliography{v2pdf}

\end{document}